# Flux through a hole from a shaken granular medium

K. Chen, M. B. Stone[†], R. Barry, M. Lohr, W. McConville, K. Klein,  B. L. Sheu, A. J. Morss[*], T. Scheidemantel, and P. Schiffer[‡]

Department of Physics and Materials Research Institute, Pennsylvania State University, University Park, Pennsylvania 16802 USA

## Abstract

We have measured the flux of grains from a hole in the bottom of a shaken container of grains.  We find that the peak velocity of the vibration, $v_{max}$, controls the flux, i.e., the flux is nearly independent of the frequency and acceleration amplitude for a given value of $v_{max}$. The flux decreases with increasing peak velocity and then becomes almost constant for the largest values of $v_{max}$.  The data at low peak velocity can be quantitatively described by a simple model, but the crossover to nearly constant flux at larger peak velocity suggests a regime in which the granular density near the container bottom is independent of the energy input to the system.

[†]Current address: Condensed Matter Sciences Division, Oak Ridge National Laboratory, Oak Ridge Tennessee 37831 USA

[*]Current address: Department of Physics, Ohio State University, Columbus, Ohio 43210 USA

[‡]e-mail: schiffer@phys.psu.edu

Granular materials such as salt or sand play an essential role in numerous industrial processes, and their intrinsically complex cooperative behavior leads to a wide range of fascinating physical phenomena.[1] Even the simple act of vertically vibrating the grains can lead to fluidization[2,3,4] and a rich variety of behavior including pattern formation,[5,6,7] heaping,[8,9] and convection.[10,11] Vibrated granular matter has thus become an important test bed for understanding the physics of granular materials, and has been characterized through a number of techniques.[12,13,14,15,16,17]

We study the flux of grains emerging though a small hole in the bottom of a container of a vibrating granular sample. This flux has analogs both in the familiar act of shaking salt from a salt shaker and also in the important statistical mechanics problem of fluid flow through a hole. While the flux has been studied extensively for unvibrated grains,[18,19,20,21] for vibrated grains the flux has only been studied in the case of flow from a vibrated hopper (in which the boundaries all slope toward an open bottom).[22,23] Our measurements complement these studies by examining the simpler case of flux from a vibrated granular medium through a hole in a flat boundary over a broader range of parameters, and we demonstrate that this quantity can be largely understood within a simple model.

Our experimental apparatus (shown schematically in the inset to Fig. 1) consisted of an open-topped cylindrical aluminum vessel (inner diameter of 152 mm) partially filled with spherical glass beads. The vessel had a flat solid bottom with a circular hole at the center through which the grains could flow. The data shown below are for bead diameter $d = 0.91 \pm 0.07$ mm unless specified otherwise, but we obtained qualitatively equivalent data for $d = 2.0 \pm 0.1$ mm and $3.0 \pm 0.1$ mm and rough sand (grain size around 1 mm). These grain sizes were large enough so that the effects of interstitial air could be neglected.[24] The hole diameter ($D$) was sufficiently large that the grains would flow even in the absence of vibration ($D/d \geq 3.6$).

The vessel was supported by four aluminum posts connected to an electromagnetic shaker controlled by two feedback accelerometers mounted on the rim of the vessel. The vessel was vibrated sinusoidally ($y_{vessel} = A\sin\omega t$) at a desired frequency, $f = \omega/2\pi$, and a peak acceleration, $\Gamma = A\omega^2/g$, which we normalize to the gravitational acceleration, $g = 9.81$ m/s$^2$. Grains flowing out from the container landed in a stationary collector suspended below the vessel. The collector was attached to an electronic balance, allowing us to record the mass of the effluent as a function of time, $m(t)$, and thus determine the time-averaged flux from the hole, $\Phi(t) = dm/dt$. The mass of glass spheres in the shaking vessel was such that there were between 20 and 90 grain layers in the container while data were acquired. The flux was found to be almost independent of mass in the shaker for the range of masses used in the experiments (variation of less than 10%). The frequencies used in our experiment ranged from 6 Hz to 400 Hz with $\Gamma$ ranging from 1.25 to 10. The range of the vibration parameters was limited by the maximum amplitude of the shaker ( $\sim$ 12 mm). The acquisition time for each measurement was 120 seconds (720 – 48000 cycles), and each of the flux values reported below was taken as an average of at least five measurements (error bars represent the standard deviation of the measurements). All data were acquired with a relative humidity of $30 \pm 10\%$, and no significant effect of humidity was observed.

Fig. 1 shows the hole diameter dependence of the "dc flux", $\Phi_0$, measured for unvibrated media. These results agree well with previous studies,[25] following the empirical equation of Beverloo *et al.*:

$$\Phi_0 = C\rho_B g^{1/2}(D - kd)^{5/2} , \qquad\qquad (1)$$

where $C$ and $k$ are dimensionless fitting parameters ($C = 0.54 \pm 0.03$, $k = 1.3 \pm 0.2$ for the best fit shown as a solid line in Fig. 1) and $\rho_B$ is the bulk density of the granular material (about 1.5 gram/cm$^3$ at rest).

Fig. 2 shows the amplitude dependence of the normalized flux under vibration, $\Phi(A)/\Phi_0$, for four different hole diameters at a constant acceleration amplitude $\Gamma = 4$. When normalized by $\Phi_0$, the data collapse to a single curve. This collapse, which is also seen for other values of $\Gamma$, implies that the Beverloo relation between flux and hole diameter (Eqn. 1) is still valid under vibration for our range of vibration parameters. This perhaps could have been anticipated, since the falling velocity of the grains from the hole ($v \sim \sqrt{2gD}$) which, multiplied with the area of the hole, accounts for the 5/2 power law in the Beverloo relation, is large compared to the typical impact velocity of particles.[14] As seen in the figure, $\Phi(A)/\Phi_0$ = 1 for small amplitudes, but, as the amplitude of vibration increases, the flux of the vibrated media decreases monotonically to less than $\Phi_0/2$. While these data may run contrary to common wisdom concerning salt shakers (i.e., that shaking with larger amplitude or acceleration will increase the flux of salt), a similar decrease is also seen in vibrated hoppers, and it can be attributed to a reduction in density associated with vibration as described in detail below. Indeed, the monotonically decreasing $\Phi(A)$ is generic to all of our data, even for smaller holes through which unvibrated flow is limited by arching, as is the case for most salt shakers.

Fig. 3 shows a series of flux data taken at different frequencies and amplitudes but a constant hole diameter of $D$ = 6.4 mm, plotted as a function of amplitude, peak velocity and acceleration. Fig. 3a shows $\Phi(A)$, and again the flux saturates at $\Phi_0$ for small amplitudes (high frequencies) and decreases for large amplitudes (low frequencies). Fig. 3b shows the same measured values of $\Phi$ plotted versus the peak velocity, $v_{max} = A\omega = \Gamma/\omega$. When plotted in this way, the data collapse onto a single characteristic curve, with the exception of a few points at low $\Gamma$ where we expect the grains to jam against the bottom for a portion of the vibration cycle. This collapse is also in sharp contrast to the separation of the data in Fig. 3c, where we plot $\Phi(\Gamma)$.

The collapse of the data in Fig. 3b suggests that $v_{max}$ (rather than $\Gamma$ or $A$) is the essential variable which determines the flux from a vibrating container, and it is notable that a similar collapse was observed in flow from a vibrated hopper. As shown in Fig. 2, vibration does not affect the hole diameter dependence of $(D-kd)^{5/2}$ in Eqn.1, thus we attribute the $v_{max}$ dependence of the flux to changes in $\rho_B$, the granular density near the vessel bottom. As grains are fluidized by vibration, the reduced volume density of grains can be described by the average gap between nearest neighbors, $s$, where $s = 0$ when the granular pile is at rest.[26]

We now offer a possible theoretical explanation for the above results, based on the above simple interpretation that the decreasing flux is due to the decreasing density of the grains near the container bottom. We follow previous authors and assume that $s$ can be described through simple energy balance between the increased potential energy of the grains and the kinetic energy imparted by collisions with the bottom:

$$\alpha m s g = m v_{max}^2 \qquad (2)$$

where $\alpha$ is a dimensionless parameter that characters the energy needed for the pile to expand $s$ from energy imparted to the granular particles by the vibrating bottom through inelastic collision. Considering the expansion between particles and ignoring geometric factors, the density under vibration then becomes:

$$\rho_B = \frac{\rho_0}{(1+\frac{s}{d})^3} , \qquad (3)$$

where $\rho_0$ is the bulk density of the particles at rest. Thus the flux under vibration can be written as:

$$\Phi = C\rho_0 g^{1/2}(D - kd)^{5/2}(1 + \frac{v_{max}^2}{\alpha g d})^{-3} \qquad (4)$$

or

$$\frac{\Phi}{\Phi_0} = (1 + \frac{v_{max}^2}{\alpha g d})^{-3} \qquad (5)$$

Under vibration, particles may experience different accelerations relative to the vessel bottom, and an effective gravity, $g_{eff}$, has been proposed to account for this.[27] When the particles are fully fluidized ($\Gamma \geq 2$), however, this effect averages out over the course of a complete vibration cycle and thus should not affect our data which are averaged over many cycles.

We now consider how well Eqn. 5 describes the data, considering only $\Gamma \geq 2$ where the grains are fluidized throughout the cycle. We immediately note that Eqn. 5 predicts that the flux should depend only on the peak velocity, $v_{max}$, in agreement with the results shown in Fig. 3. Also, as shown by the solid line in Fig. 4, the relation given in Eqn. 5 clearly fits the data for low peak velocity ($v_{max} \lesssim 0.3$ m/s) using only a single free parameter, $\alpha$, and we obtain similarly good fits for different grain diameters ($d = 2$ and 3 mm). The data on different grains suggest that $\alpha$ is proportional to $d$ (Fig. 4 inset). This indicates that greater velocity is needed to separate larger particles, a supposition which seems physically reasonable.

At the largest values of $v_{max}$, the normalized flux as plotted in Fig. 4 deviates significantly from the behavior expected from Eqn. 5, and is almost constant for roughly a factor of two in $v_{max}$. The deviation is not surprising, however, since Eqn. 2 was derived under the condition that the particle separation is much smaller than the particle diameter, $s \ll d$, and the deviation occurs when $s/d \sim 0.4$. The roughly constant value of $\Phi(v_{max})$ for the largest $v_{max}$ suggests that the density near the bottom of the vessel is approximately independent of vibration intensity for large amplitude vibrations. This is consistent with previous findings that the total expansion of a strongly vibrated medium does not continue to increase with vibration intensity.[28,29] We speculate that this behavior is rooted in the inelastic nature of granular collisions. When the vibration rate and granular temperature increase, the collision rate will increase,[26] and the restitution coefficient will decrease.[30,31] Both the collision rate increase and the restitution coefficient decrease may serve to balance the increased energy input from increasing vibration amplitude, keeping the granular temperature near the bottom from increasing and resulting in a constant density. The local temperature may also contribute to this constant flux when the fluidized grain velocities become comparable to the falling velocity which enters the Beverloo relation for the flux of grains from a static container. Future experiments, such as detailed optical or MRI measurements of granular dynamics," may be able to probe these possibilities further.

While the above analysis is somewhat simplistic in that we assume the flux is controlled by the density near the container bottom and also the energy balance relationship stated in Equation 2, the fit to the data is quite reasonable over a broad range of vibration intensities. The collapse of the data for different hole sizes and the quality of the fit both indicate that measurements of the flux from a vibrationally excited granular material can be added to the growing toolbox of experimental techniques for the understanding of this important model system."[32]

**Acknowledgements**

Research was supported by the NASA through grant NAG3-2384 and the NSF REU program through grant DMR 0305238.

**Figure Captions**

**Figure 1.** (Color online) The measured flux in the absence of vibration ($\Phi_0$) as a function of hole diameter ($D$). The solid line is a fit to Eqn. 1 for $d = 0.91$ mm (symbol size is greater than the error bar). Inset: A schematic of the experimental apparatus.

**Figure 2.** (Color online) The granular flux as a function of vibration amplitude, $A$, for acceleration of $\Gamma = 4$, frequency $f = 10$ - 400 Hz, and particle diameter $d = 0.91$ mm (error bars come from the standard deviation in the results of multiple runs). The flux data for different hole sizes are normalized by corresponding $\Phi_0$. Note that the data for different hole sizes all collapse onto a single curve (a similar collapse is also seen for other values of $\Gamma$).

**Figure 3.** (Color online) The granular flux under vibration as a function of **a.** vibration amplitude ($A$), **b.** peak velocity ($v_{max}$) and **c.** acceleration ($\Gamma$) where the error bars come from the standard deviation in the results of multiple runs (particle diameter $d = 0.91$ mm). The amplitude range for different accelerations: $\Gamma = 1.25$, $A = 1.9 \times 10^{-3} - 3.1$ mm; $\Gamma = 1.5$, $A = 2.3 \times 10^{-3} - 10$ mm; $\Gamma = 2$, $A = 3.1 \times 10^{-3} - 8.8$ mm; $\Gamma = 4$, $A = 6.2 \times 10^{-3} - 9.9$ mm; $\Gamma = 6$, $A = 9.3 \times 10^{-3} - 6.6$ mm and $\Gamma = 10$, $A = 1.6 \times 10^{-2} - 11$ mm.

**Figure 4.** (Color online) A comparison of the peak velocity dependence of the flux, $\Phi(v_{max})$, with the model described in the text (particle diameter $d = 0.91$ mm). The solid line is a fit of the data from Fig. 3 to Eqn. 5. Data with $\Gamma < 2$ are not included due to the fluidization threshold. A good agreement of data and theory is obtained for small peak velocities, but a clear break-down of the agreement is seen at the largest peak velocities. Inset: fitting

parameter $\alpha$ as a function of $d$ for $d$ = 0.91 mm, 2 mm, and 3 mm, symbol size is greater than the error bar, the slope of the fitted straight line is $23.4 \pm 1.0$.

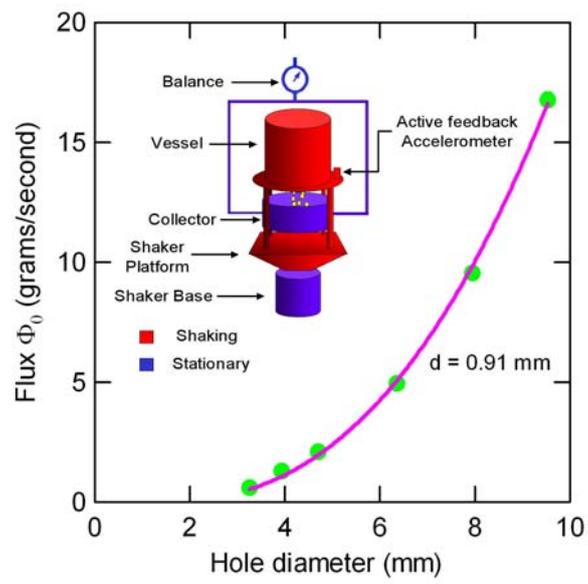

Figure 1 Chen *et al*

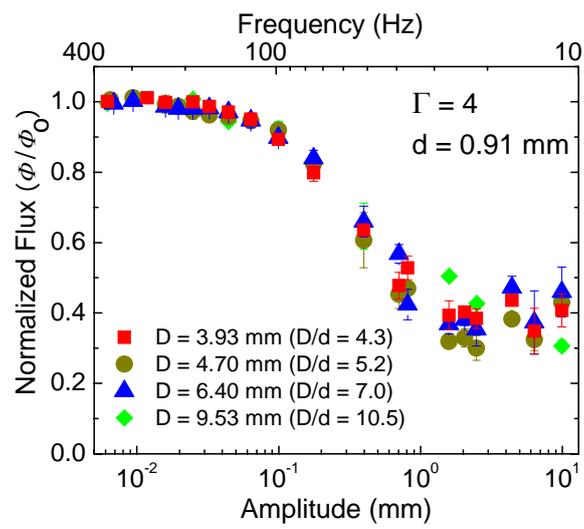

Figure 2  Chen *et al.*

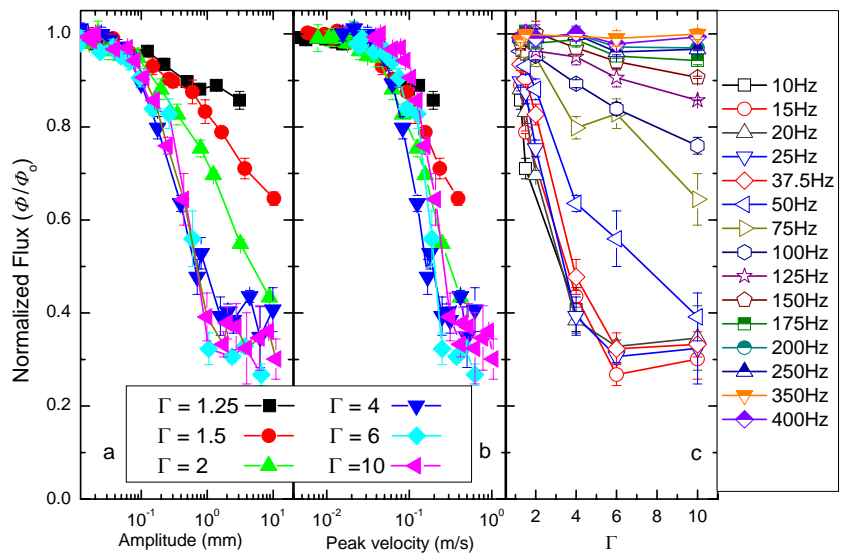

Figure 3 Chen *et al*

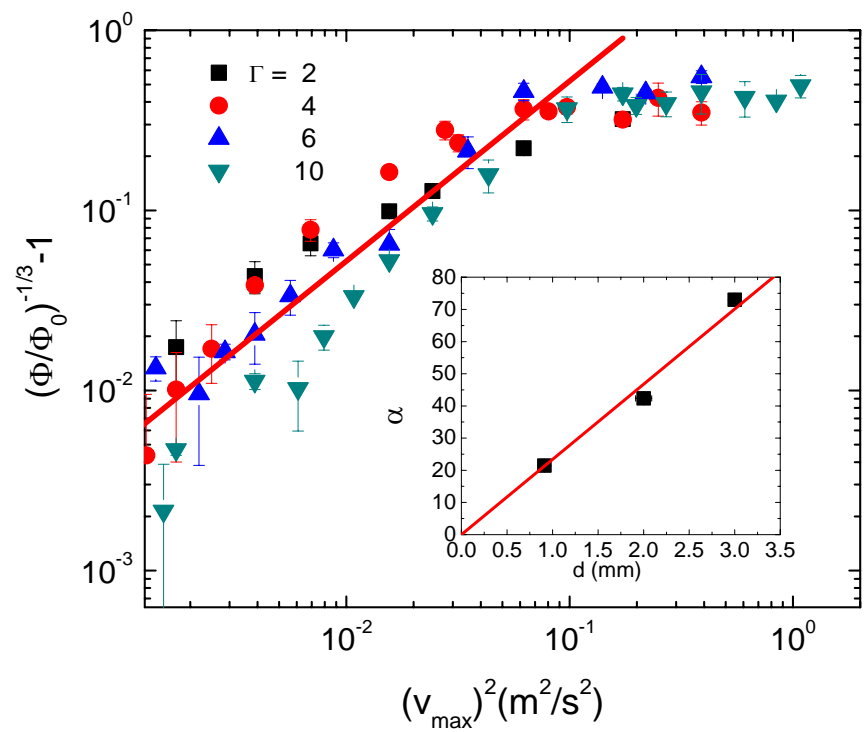

Figure 4  Chen *et al.*